\documentclass[%
 reprint,
superscriptaddress,
 amsmath,amssymb,
 aps,
]{revtex4-2}

\usepackage{graphicx}% Include figure files
\usepackage{dcolumn}% Align table columns on decimal point
\usepackage{bm}% bold math
\usepackage[hidelinks]{hyperref}

\usepackage{xcolor}

\begin{document}

\preprint{APS/123-QED}

\title{Sub-Yield Dynamics in Yield-Stress Materials}

\author{Alice Woodbridge}
\thanks{These authors contributed equally to this work.}
\affiliation{Department of Physics and Astronomy and Manchester Centre for Nonlinear Dynamics, The University of Manchester, Oxford Road, Manchester M13~9PL, United Kingdom}

\author{Kasra Amini}
\thanks{These authors contributed equally to this work.}
\affiliation{FLOW and Fluid Physics Laboratory, Department of Engineering Mechanics, KTH, Stockholm SE-100 44, Sweden}

\author{Fredrik Lundell}
\affiliation{FLOW and Fluid Physics Laboratory, Department of Engineering Mechanics, KTH, Stockholm SE-100 44, Sweden}

\author{Outi Tammisola}
\affiliation{FLOW and SeRC (Swedish e-Science Research Centre), Department of Engineering Mechanics, KTH, Stockholm SE-100 44, Sweden}

\author{Anne Juel}
\affiliation{Department of Physics and Astronomy and Manchester Centre for Nonlinear Dynamics, The University of Manchester, Oxford Road, Manchester M13~9PL, United Kingdom}

\author{Robert J. Poole}
\affiliation{School of Engineering, University of Liverpool, Liverpool L69~3GH, United Kingdom}

\author{Cl\'{a}udio P. Fonte}
\thanks{Contact author: claudio.fonte@manchester.ac.uk}
\affiliation{Department of Chemical Engineering, The University of Manchester, Oxford Road, Manchester M13~9PL, United Kingdom}

\date{\today}% It is always \today, today,

\begin{abstract}
The mechanical response of yield-stress materials below the yield point remains a subject of debate. Two of the most widely used constitutive models for these materials offer fundamentally conflicting views: one permits plastic flow at all stress levels, the other assumes entirely recoverable viscoelasticity below yield. Using parallel superposition rheometry, we test the sub-yield behaviour of a microgel and an emulsion. When residual slip effects are properly accounted for, both fluids exhibit bounded, periodic strain responses, offering compelling evidence that they do not flow in the studied regime. Our results indicate that the sub-yield regime is underpinned by nonlinear viscoelasticity and underscore the need for improved constitutive relations that capture such effects without treating yielding as a precursor for nonlinearity.
\end{abstract}

\maketitle

%\tableofcontents

Yield-stress materials (YSMs) are among the most common and versatile substances in everyday life, found in applications ranging from food and cosmetics to concrete. They are relevant across strikingly different length scales -- from geophysical flows such as mudslides and lava~\cite{Balmforth2014} to the microscale with self-assembled peptide gels for 3D cell culture and controlled drug delivery~\cite{gao_controlling_2017}. These materials share a defining feature: a transition from solid- to fluid-like behaviour when the applied stress exceeds a critical yield threshold. Many YSMs also exhibit an elastoviscoplastic response that combines elastic, viscous, and plastic characteristics, depending on the applied stress and timescale~\cite{Abdelgawad2024}: elastic response entails reversible deformation with stress proportional to strain; viscous response entails rate-dependent dissipation with stress proportional to strain rate; plastic (viscoplastic) behaviour denotes irreversible deformation associated with yielding, i.e. the onset of sustained flow and unrecoverable strain once a yield threshold is exceeded~\cite{Bonn2017}. Despite this common phenomenology, their microstructures vary widely, encompassing soft-bead microgels, emulsions, and dense suspensions. Notwithstanding their practical significance and study for over a century~\cite{Bingham1916}, the physical understanding of the processes leading to yielding remains incomplete, even for non-thixotropic fluids~\cite{Bonn2017}, and there is persistent ambiguity in the identification and definition of ``yielding"~\cite{BARNES1999133, DINKGREVE2016233, Coussot2014}.

To illustrate the diversity of views surrounding the physical processes that lead to yielding, we consider two widely adopted, yet fundamentally different, approaches to describing the behaviour of YSMs: the Saramito–Herschel–Bulkley (SHB) model~\cite{SARAMITO2007,SARAMITO2009} and the Kamani–Donley–Rogers (KDR) model~\cite{RogersPRL2021}. In the SHB model, yielding is treated as a sharp, stress-based transition: below the yield stress, the material behaves as a linear viscoelastic solid, while past yield it exhibits plastic flow according to the Herschel–Bulkley relation, combined with viscoelastic behaviour. No plastic flow occurs below yield, and deformation is entirely recoverable in this region. In contrast, the KDR model does not impose a strict yield criterion despite using the material's yield stress as a model parameter. Instead, it introduces a strain-rate-dependent mobility function that allows for plastic flow at all stress levels, resulting in a continuous and smooth transition between solid- and fluid-like regimes~\cite{RogersPRL2021, Griebler2024}.
This smooth crossover enables the KDR model to more effectively capture the response of YSMs in standard large-amplitude oscillatory shear tests (such as strain amplitude sweeps) compared to models with abrupt yielding.

A recent experimental study on the fluidisation of sessile drops of YSMs under vertical sinusoidal oscillations appears to cast doubt on the notion of plastic flow occurring prior to yielding \cite{GARG2021104595, GARG2022100067}. 
In these experiments, the drop exhibits periodic viscoelastic oscillations without measurable spreading, suggesting the absence of unrecoverable deformation below the spreading threshold. However, owing to the inherently protorheological \cite{Hossain2024} nature of the oscillated drop setup, spreading could not be mapped directly to the yield stress.
Nonetheless, these observations suggest that combining steady gravitational stress (from the drop’s weight) with sinusoidal stress (from the oscillating platform), which together induce asymmetric, time-dependent deformation, may help identify the physical mechanism underlying the onset of unrecoverable deformation. However, a well-controlled analogous rheometric investigation is needed to distinguish whether unrecoverable deformation arises globally from plastic flow at any stress level -- as permitted by the KDR model -- or from localised fluidisation in regions where the stress exceeds the yield threshold, while other parts of the drop remain in a recoverable solid-like state -- as described by the SHB model.
Crucially, it raises a central and longstanding question: \emph{Do yield-stress materials flow prior to yielding?} For clarity, by \emph{flow} we mean pre-yield irreversible deformation attributable to the bulk response, rather than any long-term behaviour associated with material ageing, which the SHB and KDR models considered here are not designed to capture.
 
In this Letter, we address the question using parallel superposition shear rheometry \cite{VERMANT1998}, wherein an oscillatory stress is superimposed on a steady offset so that the total forcing remains sub-yield throughout. Stress-controlled tests, rather than strain control, avoid enforcing deformations that locally may exceed the yield threshold near the yield point \cite{Korculanin2017}. To the best of our knowledge, this is the first study to explore experimentally the sub-yield response of elastoviscoplastic materials such as gels and emulsions using parallel superposition rheometry. We note the seminal work of \citeauthor{Pons2015}~\cite{Pons2015} on ``secular drift'' in granular (glass-bead) packings under similar bias-plus-oscillation protocols, though their results concern a different material class and flow geometry. Although our focus is the sub-yield response, we also include measurements beyond yield to delineate the transition -- showing that sub-yield behaviour persists up to the yield point and breaks down above it.

%%% Theoretical analysis

We start by theoretically analysing the SHB and KDR model responses to a shear stress $\sigma_{xy} = \left[\sigma_0 + \epsilon \sin(\omega t)\right] u(t)$ composed of a sinusoidal component with amplitude $\epsilon$ and frequency $\omega$ superimposed on a constant stress offset $\sigma_0$, applied from rest via the Heaviside step function $u(t)$. The SHB model in one-dimensional shear flow is given by
\begin{equation}
        \begin{cases}
          \dfrac{1}{G}\left(\dot{\tau}_{xx}-2\dot{\gamma}\,\tau_{xy} \right)+\max \left(0, \dfrac{\tau_D - \tau_Y}{k \tau_D^n}\right)^{1/n} \tau_{xx}=0 \\
          \dfrac{\dot{\tau}_{xy}}{G}+\max \left(0, \dfrac{\tau_D -\tau_Y}{k \tau_D^n}\right)^{1/n} \tau_{xy}=\dot{\gamma}
        \end{cases},
        \label{eqn:saramito}
\end{equation}
where $\tau_{xy}$ and $\tau_{xx}$ are the shear and normal components of the extra stress tensor, respectively, and $\dot{\gamma}$ is the strain rate. The yield stress, $\tau_Y$, elastic modulus, $G$, and consistency and flow indices, $k$ and $n$, respectively, are the model parameters. The term $\tau_D = (\tau_{xx}^2/4 + \tau_{xy}^2)^{1/2}$ represents the magnitude of the deviatoric component of the extra stress, which sets fluidisation according to the von Mises criterion ($\tau_D > \tau_Y$). The total shear stress, including the solvent contribution, is given by $\sigma_{xy}=\tau_{xy}+\eta_s \dot{\gamma}$, where $\eta_s$ is a solvent viscosity. Below yielding the SHB material behaves like a Kelvin-Voigt solid. Under these conditions, \eqref{eqn:saramito} admits an analytical solution for the strain,
\begin{multline}
\gamma = \frac{\sigma_0}{G} 
+ \frac{\epsilon G}{\eta_s^2 \omega^2 + G^{2}} \sin(\omega t) 
- \frac{\epsilon \eta_s \omega}{\eta_s^2 \omega^2 + G^{2}} \cos(\omega t) + \\
+ e^{-\frac{Gt}{\eta_s}} 
\left( \frac{\epsilon \eta_s \omega}{\eta_s^2 \omega^2 + G^{2}} 
- \frac{\sigma_0}{G} \right),
\label{eqn:solution_saramito}
\end{multline}
for an initially strain-free material. Once the yield stress is exceeded, the material transitions into an elastoviscoplastic flow regime, and \eqref{eqn:saramito} must be solved numerically.  

\begin{figure}[t!]
    \centering
    \includegraphics[scale=1.0]{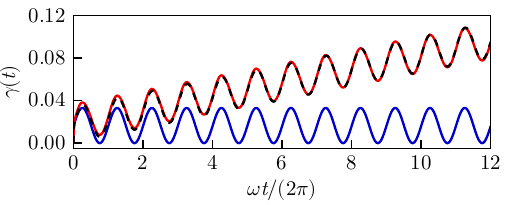}
    \caption{Comparison between the predicted strain responses of the SHB model (solid blue line), the numerical solution of the KDR model (solid red line), and the analytical approximation of the KDR model (dashed black line) under parallel superposition of steady and oscillatory stress. Constitutive parameters correspond to a 2~g/L Carbopol gel (see Table~S1 in the \emph{Supplemental Material}~\cite{SM}). The applied stress components are $\sigma_0 = \epsilon = 0.1\tau_Y$, and $\omega = 1$~rad/s.}
    \label{fig:theoryComparison}
\end{figure}

% Added here to get the correct numbering for this reference from the SM
\nocite{dinkgreve2018carbopolThixo}

The KDR constitutive model for one-dimensional shear flow is given by
\begin{equation}
      \sigma_{xy} + \lambda(\dot{\gamma})\, \dot{\sigma}_{xy}= \left(\frac{\tau_Y}{|\dot{\gamma}|} +k |\dot{\gamma}|^{n-1} \right)\left( \dot{\gamma}+\frac{\eta_s}{G} \ddot{\gamma}\right),
      \label{eqn:kdr}
\end{equation}
where $|\dot{\gamma}|$ is the absolute value of the shear rate, and the model parameters that are common to the SHB model have the same meaning apart from $\eta_s$ that here represents a structural viscosity. The parameter $\lambda$ is a rate-dependent relaxation time given by
\begin{equation*}
    \lambda(\dot{\gamma}) = \left(\frac{\tau_Y}{|\dot{\gamma}|} +k |\dot{\gamma}|^{n-1} +\eta_s\right) / G.
\end{equation*}
In the limit of small shear rate amplitudes, where the condition
$\tau_Y /|\dot{\gamma}| \gg k |\dot{\gamma}|^{n-1} + \eta_s$ is satisfied, \eqref{eqn:kdr} becomes
\begin{equation}
\eta_s \ddot{\gamma} + G \dot{\gamma} = \dot{\sigma}_{xy} + \left| \dot{\gamma} \right| \frac{G}{\tau_Y} \sigma_{xy}.
\label{eqn:kdr_simplified}
\end{equation}
To approximate the solution of \eqref{eqn:kdr_simplified} analytically, we treat the nonlinear term as a small perturbation to the linear Maxwell problem. In this regime, the response is dominated by a single harmonic, so $\dot{\gamma}(t) \approx A \omega \cos(\omega t + \varphi)$, where
\begin{equation*}
A = \frac{\epsilon}{\sqrt{\eta_s^2 \omega^2 + G^2}}, \quad \text{and} \quad \varphi = \arctan\left(-\frac{\eta_s \omega}{G}\right).
\end{equation*}
Finally, $\left|\dot{\gamma}\right| \approx A\omega\left|\cos(\omega t + \varphi)\right| \approx 2 A\omega/\pi$, when retaining only the fundamental term of the Fourier series expansion. For an initially strain-free stationary material, we then obtain
\begin{equation}
      \gamma =\Pi_0+\Pi_1 e^\frac{-Gt}{\eta_s}+\Pi_2 \sin(\omega t) + \Pi_3 \cos(\omega t) + \Pi_4 t,
      \label{eqn: approx solution kdr}
\end{equation}
where
\begin{equation*}
    \Pi_0 = \dfrac{\xi \epsilon}{\omega} + \dfrac{\sigma_0 (1 - \xi \eta_s)}{G},
\end{equation*}
\begin{equation*}
    \Pi_1 = (1-\xi \eta_s)\left( \dfrac{\epsilon \eta_s \omega}{\eta_s^2 \omega^2 + G^2} - \dfrac{\sigma_0}{G} \right),
\end{equation*}
\begin{equation*}
    \Pi_2 = \dfrac{\epsilon G(1-\xi \eta_s)}{\eta_s^2 \omega^2 + G^2}, \quad \Pi_3 =  - \dfrac{\epsilon (\eta_s \omega^2 + \xi G^2)}{\omega (\eta_s^2 \omega^2 + G^2)},
\end{equation*}
and $\Pi_4 = \xi \sigma_0$ for $\xi \equiv 2 A \omega/(\pi \tau_Y)$. Augmenting the Fourier representation of $\lvert \dot{\gamma} \rvert$ by one additional mode would yield only perturbative corrections to $ \Pi_2 $ and $ \Pi_3 $ and introduce a small filtered second-harmonic ($2\omega$) component into the response; to the order retained, all other terms remain unchanged. The term $\Pi_4 t$ in \eqref{eqn: approx solution kdr} introduces a linear time-dependent contribution to the shear strain, which is absent in the SHB model solution [cf.~\eqref{eqn:solution_saramito}]. Note that the magnitude of this additional term is governed by the parameters associated with all three components of the imposed shear stress protocol ($\sigma_0$, $\epsilon$, and $\omega$) as
\begin{equation}
    \Pi_4 t = \frac{2}{\pi}  \frac{\sigma_0 \epsilon}{\tau_Y\sqrt{\eta_s^2 + (G/\omega)^2}} t,
    \label{eq:linear_term_kdr}
\end{equation}
revealing its dependence on the product $\sigma_0\epsilon$. Hence, this linear-in-time contribution persists as long as both the steady and oscillatory stress components are non-zero (i.e., $\sigma_0 \epsilon \ne 0$). 

Fig.~\ref{fig:theoryComparison} shows the close agreement between the numerical solution of the KDR model and its approximate analytical solution for typical strain amplitude values applied in the parallel superposition tests, thus validating \eqref{eqn: approx solution kdr} in the regime of interest. This figure also highlights the key distinction between the behaviours of the SHB and KDR models under sub-yielding stress conditions and indicates that this should be measurable experimentally. While the SHB model evolves towards a bounded periodic strain with an offset of $\sigma_0/G$, the KDR model instead predicts a strain response characterised by linear growth superimposed with the additional sinusoidal and constant terms in \eqref{eqn: approx solution kdr}. From the approximate analytical solution of the KDR model, it follows that when either component of the imposed stress signal is set to zero -- whether under a pure step stress or a purely oscillatory stress input -- the strain responses predicted by the SHB and KDR models become similar. This similarity hinders the ability to discriminate between models and, consequently, to identify the most appropriate framework for describing sub-yield material behaviour. This highlights the importance of using complex flow protocols like the one adopted in this work to probe the rheological behaviour of structured fluids.

\begin{figure}[t!]
    \centering
    \includegraphics[scale=1.0]{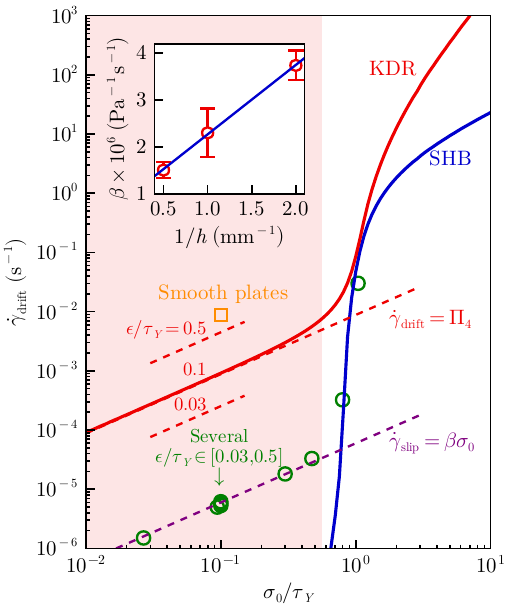}
    \caption{Linear strain drift in the parallel superposition measurements in a Carbopol gel. Main: Strain rate drift from experiments (green markers), SHB model (blue solid), KDR model (red solid); red dashed: analytical KDR approximation [Eq.\eqref{eq:linear_term_kdr}]; purple dashed: slip-only term, $\dot{\gamma}_\mathrm{drift}=\dot{\gamma}_\mathrm{slip}=\sigma_0\beta$. At $\sigma_0=0.1\tau_Y$, four overlapping markers correspond to $\epsilon/\tau_Y \in \{0.03,0.1,0.3,0.5\}$; all other data and model curves use $\epsilon=0.1\tau_Y$. The orange square shows the drift measured with smooth plates at $\sigma_0=\epsilon=0.1\tau_Y$. The shaded region marks the fully unyielded limit in the parallel-disk geometry, $\sigma_0+\epsilon\le 2\tau_Y/3$. Inset: phenomenological slip parameter (red circles) versus gap between crosshatched plates at $\sigma_0=\epsilon=1 \, \mathrm{Pa}$ and $\omega=1~\mathrm{rad \, s^{-1}}$; error bars denote the standard error; blue solid: best linear fit.}
    \label{fig:slip_factor}
\end{figure}

We now analyse the experimental strain responses obtained via parallel superposition rheometry for two representative YSMs: a Carbopol gel and a commercial body lotion. Even with cross-hatched plates -- used to minimise wall slip -- the measured strain response showed a small linear drift component, $\gamma \sim \dot{\gamma}_\mathrm{drift}t$. Fig~\ref{fig:slip_factor} compares the experimental $\dot{\gamma}_\mathrm{drift}$ with that predicted by the KDR model, where the full numerical solution and the term $\Pi_4$ from its analytical approximation [cf.~\eqref{eq:linear_term_kdr}] are virtually indistinguishable below yield (see Fig.~\ref{fig:theoryComparison}). For reference, Fig.~\ref{fig:slip_factor} also shows the drift measured with smooth plates at $\sigma_0=\epsilon=0.1\,\tau_Y$, which is about three orders of magnitude larger than with crosshatched plates under identical conditions. Unlike the strain drift predicted by the KDR model, the rate of linear drift in the experiments is sensitive to the gap height between the rheometer plates and scales with the steady component of the applied stress, while remaining insensitive to the oscillatory amplitude. Moreover, its magnitude is two orders of magnitude smaller than the drift predicted by the KDR model. Non-local fluidity effects are known to appear in confined geometries when the gap size is comparable to the cooperativity length (tens of micrometres for emulsions and Carbopol gels) \cite{Goyon2008, Bocquet2009, Vleminckx2020}. However, the gaps between the parallel disks used in this work are two to three orders of magnitude larger, where such effects are negligible. Moreover, non-local effects on fluidity -- or, more generally, any effect linked to bulk properties -- would imply a sensitivity to oscillatory amplitude at fixed mean stress, i.e. varying $\epsilon$ at fixed $\sigma_0$ would alter the effective fluidity and thus the rate of linear drift. In our measurements, the magnitude of the drift shows strong gap and roughness dependence but no systematic dependence on $\epsilon$ over a wide range of imposed stresses (cf. Fig~\ref{fig:slip_factor} for $\sigma_0/\tau_Y=0.1$), supporting instead an interfacial (wall slip) interpretation \cite{Zhang2017}.

To account for the effect of residual slip and isolate the bulk material behaviour, we used a Navier slip model to subtract the slip-induced strain from the parallel superposition measurements (see Appendix~\ref{App:Slip} in the End Matter for details). Fig.~\ref{fig:slip_factor} shows the close agreement between the adopted slip model across a wide range of stresses up to yielding conditions. Above yield, slip is negligible and the drift rate collapses onto the SHB model prediction. These findings underscore the role of residual wall slip as a possible cause of result misinterpretation, particularly in the sub-yield regime where apparent unrecoverable deformation may in fact arise from these interfacial effects rather than true bulk flow. This makes slip a critical experimental challenge in parallel superposition rheometry and related techniques, such as large- and small-amplitude oscillatory shear.

\begin{figure}[t!]
    \centering
    \includegraphics[scale=1.0]{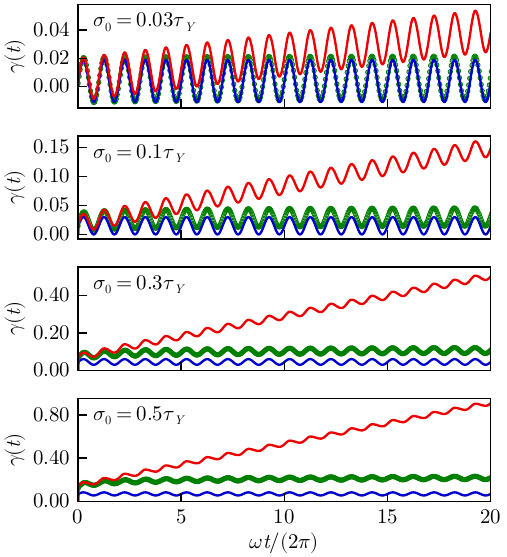}
    \caption{Comparison of parallel-superposition rheometry data (green circles) with full numerical predictions from the SHB (blue) and KDR (red) models over the first 20 cycles for Carbopol. The imposed stress is $\sigma(t)=\sigma_{0}+\epsilon\sin(\omega t)$, with $\sigma_{0}/\tau_Y\in \{0.03,\,0.1,\,0.3,\,0.5\}$ (top to bottom), $\epsilon=0.1 \tau_Y$ and $\omega=1~\mathrm{rad \, s^{-1}}$.}
    \label{fig:Exp_SHB_KDR}
\end{figure}

After subtraction of residual slip effects, the experimental data in Fig.~\ref{fig:Exp_SHB_KDR} for Carbopol gel show a material response consisting of pure sinusoidal deformation superimposed on a constant strain offset. This response is consistent with that of a viscoelastic solid: deformation remains bounded and fully recoverable, with no indication of continuous flow. Notably, this behaviour persists throughout the experiment; see Sec.~S2 of the \emph{Supplemental Material}~\cite{SM} for the full 75 cycles. Similar results have been obtained for the body lotion (see Fig.~\ref{fig:Exp_SHB_KDR_bodylotion} in End Matter).

We compare the experimental data in Fig.~\ref{fig:Exp_SHB_KDR} with the numerical solutions of the SHB and KDR models. Because parallel-disk rheometry controls torque rather than setting a uniform shear stress field, the local stresses vary radially; we account for this exactly (see Appendix~\ref{App:Plate-plate} in the End Matter) and obtain model predictions by integrating the SHB and KDR constitutive responses over the radial position. Model responses were generated using parameters independently determined from steady and oscillatory shear tests, without additional fitting to parallel superposition rheometry data. A summary of the key rheological parameters extracted from these tests is provided in Table~S1 in the \emph{Supplemental Material}. Full experimental details of material preparation and characterisation, and the parallel superposition rheometry protocol used are provided in Sec.~S1 of the \emph{Supplemental Material}~\cite{SM} and Appendix~\ref{App:PSP} in the End Matter, respectively.

The SHB model is able to systematically capture the solid-like material behaviour observed experimentally for both materials (see Fig.~\ref{fig:Exp_SHB_KDR} for Carbopol and Fig.~\ref{fig:Exp_SHB_KDR_bodylotion} in End Matter for the body lotion). The model’s predictions show very good agreement with both the amplitude and phase of the periodic steady-state strain response. However, while the SHB model captures the strain offset reasonably well at lower stress magnitudes, deviations from experimental data become more pronounced as the applied stress increases, particularly for the Carbopol gel. Crucially, however, the absence of any secular drift in the experiments contrasts with the KDR prediction under the same loading: Fig.~\ref{fig:Exp_SHB_KDR} shows that the KDR model does not capture the sub-yield behaviour of Carbopol (or the emulsion -- see Fig.~\ref{fig:Exp_SHB_KDR_bodylotion} in End Matter), predicting instead a pronounced linear growth in strain that is not observed experimentally -- even well below the yield stress. This linear drift reflects liquid-like, unrecoverable deformation and, as shown theoretically, arises only under parallel superposition of steady and oscillatory stresses, vanishing when either component is removed. This is consistent with the good performance of the KDR model in standard rheometric protocols such as simple steady and oscillatory shear \cite{RogersPRL2021, Griebler2024}. However, its inability to capture the bounded strain response under parallel superposition of stresses highlights a fundamental limitation in its formulation. Moreover, while a recent Letter supported by the KDR model \cite{Keane2025} proposes that the physics underlying linear viscoelasticity is preserved throughout yielding, our experimental results provide opposing evidence that pronounced nonlinear viscoelastic behaviour emerges even in the absence of fluidisation. Accordingly, the hypothesis put forward in Ref.~\cite{Keane2025} that yielding is universally governed by the non-local mechanisms encoded in the KDR model, in which elastic recoverable properties enhance plastic unrecoverable behaviours, does not appear to hold under the rheometric protocol employed in this study. For clarity, we use the term “nonlinear” here not in the classical sense of waveform distortion -- i.e., a signal that can no longer be described by a single Fourier mode -- but to indicate \emph{constitutive nonlinearity} \cite{Fu_Ogden_2001}, where the material’s elastic and viscous properties depend on the applied stress magnitude, even though the strain response remains a clean sine wave at the input frequency, as evidenced by the results presented in Fig.~\ref{fig:Exp_SHB_KDR}.

\begin{figure}[t!]
    \centering
    \includegraphics[scale=1.0]{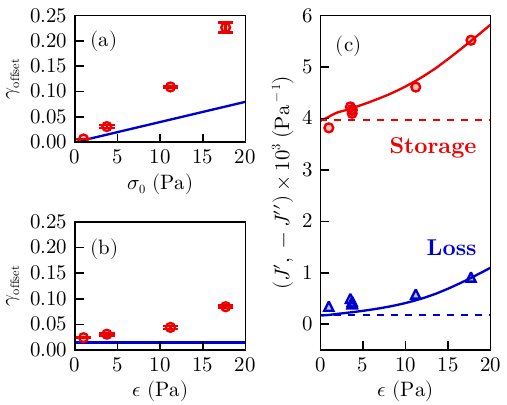}
    \caption{ (a, b) Comparison between experimental strain offsets (red circles) and predictions from the SHB model (blue line) in parallel superposition tests on Carbopol gel. Error bars indicate the standard error across independent repetitions. (a) fixed oscillatory amplitude $\epsilon = 0.1\tau_Y$; (b) fixed mean stress $\sigma_0 = 0.1\tau_Y$. (c) Storage (red) and loss (blue) compliances as functions of $\epsilon$ for Carbopol gel. Markers denote experimental values from parallel superposition rheometry; solid lines correspond to conventional oscillatory shear measurements; dashed lines show predictions from the SHB model. At $\epsilon=0.1 \tau_Y$ there are four overlapping markers for values of $\sigma_0 = [0.03, \, 0.1, \, 0.3, \, 0.5]\tau_Y$. The remaining points were measured for $\sigma_0 = 0.1\tau_Y$. For all measurements $\omega=1~\mathrm{rad \, s^{-1}}$.}
    \label{fig:OffsetCompliance}
\end{figure}

The strain offset, defined as the time-average of the strain over a cycle, $\gamma_\mathrm{offset} = \omega/(2\pi)\oint \gamma(t) \, \mathrm{d}t$, is examined in more detail in Fig.~\ref{fig:OffsetCompliance}(a, b) for Carbopol as a function of both the mean stress $\sigma_0$ and the oscillatory amplitude $\epsilon$ [see Fig.~\ref{fig:OffsetCompliance_bodylotion}(a, b) in the End Matter for the body lotion].
The emergence of pronounced deviations from $\gamma_\mathrm{offset}=\sigma_0/G$ as both $\sigma_0$ and $\epsilon$ increase is consistent with the presence of a stress-dependent elastic response, not captured by the SHB model, which assumes fixed material properties below yield. These results suggest that further refinements -- such as incorporating stress- or strain-dependent material properties -- may be needed to more fully capture the complexity of sub-yield behaviour.

To quantify the oscillatory response, we calculate the storage, $J^{\prime} = \omega/(\epsilon \pi) \oint \gamma(t) \sin(\omega t) \, \mathrm{d}t$, and loss, $J^{\prime\prime} = \omega/(\epsilon \pi) \oint \gamma(t) \cos(\omega t) \, \mathrm{d}t$, compliances of both materials. Fig.~\ref{fig:OffsetCompliance}(c) presents the compliances for Carbopol as a function of the amplitude of the oscillatory component of the applied stress signal $\epsilon$. The figure also includes the values predicted by the SHB model [cf.~\eqref{eqn:solution_saramito}] and the ones calculated from the storage, $G^\prime$, and loss, $G^{\prime\prime}$, moduli obtained from oscillatory shear tests (see Sec.~S1 in the \emph{Supplemental Material}~\cite{SM} for details). The dynamic moduli are converted into compliances using the expressions $J^\prime = G^\prime / (G^{\prime 2} + G^{\prime\prime 2})$ and $J^{\prime\prime} = -G^{\prime\prime} / (G^{\prime 2} + G^{\prime\prime 2})$ \cite{Morrison2001}. For Carbopol gel, the experimental data and SHB model predictions in Fig.~\ref{fig:OffsetCompliance}(c) show good agreement at low values of $\epsilon$, but begin to diverge as the stress amplitude increases. Notably, the experimental compliances increase with $\epsilon$, while the SHB model predicts a constant compliance across the entire range. This discrepancy reinforces the conclusion that the material's sub-yield behaviour cannot be adequately captured by a linear viscoelastic framework. The results obtained from oscillatory shear tests also support this interpretation, showing a similar increase in compliance with $\epsilon$. In addition, the compliances appear largely insensitive to the mean stress $\sigma_0$, suggesting that the oscillatory response is primarily governed by the amplitude of the dynamic component $\epsilon$ [see overlapping markers at $\epsilon=4$~Pa in Fig.~\ref{fig:OffsetCompliance}(c)]. The experimental and SHB-predicted compliances for the body lotion [Fig.~\ref{fig:OffsetCompliance_bodylotion}(c) in End Matter] show closer agreement across the entire range of tested oscillatory stresses. This agreement is likely due to the comparatively lower values of $\sigma_0$ and $\epsilon$ used in these tests: because the emulsion’s yield stress is markedly lower than Carbopol’s, it can only be probed over a narrower stress window, limiting access to the large-amplitude regime where strong nonlinearity drives deviations from the SHB model.
As with the Carbopol gel, the compliance values do not exhibit a significant dependence on the mean stress $\sigma_0$. Interestingly, the compliances derived from the dynamic moduli for the body lotion show a non-constant dependence on $\epsilon$, indicating the onset of nonlinearity even at relatively small stress amplitudes. This suggests that strain-controlled tests may impose deformation modes that do not arise naturally under stress control, where, in the latter, the material can respond more freely and the onset of nonlinearity may be suppressed.

Our findings also support the results from the vertically vibrated drop experiments that originally motivated this work~\cite{GARG2021104595}. In these experiments, the effective storage modulus was observed to decrease monotonically with increasing forcing amplitude, even prior to the onset of noticeable spreading. The results presented here, obtained in a rheometric flow, suggest that this behaviour reflects intrinsic constitutive nonlinearity. 

In conclusion, we use parallel superposition rheometry to provide compelling evidence that YSMs do not deform irreversibly (flow) before they yield, thereby addressing a longstanding question. At the same time, they also exhibit complex, stress-dependent recoverable deformation. Our results support a sub-yield response governed by non-Hookean viscoelastic solid-like material behaviour, pointing to the need for improved constitutive models that incorporate such effects without relying on yielding as a precondition for the onset of nonlinearity. The emergence of this behaviour across materials with distinct microstructures highlights its generality, as evidenced by their occurrence in both gels and emulsions. We anticipate that these findings will guide the development of next-generation models for the accurate description of sub-yield mechanics in structured fluids, with broad relevance across science and engineering.

\textit{Acknowledgements}---A.W. acknowledges support from The University of Manchester through the Manchester Mathematical Modelling in Science \& Industry initiative. K.A. acknowledges funding from European Union’s Horizon~2020 research and innovation programme under the Marie Skłodowska-Curie grant agreement No.~955605 YIELDGAP.

\textit{Data Availability}---The data that support the findings of this Letter are openly available at~\cite{rheoData}.

% \nocite{*}

\bibliography{bibliography}% Produces the bibliography via BibTeX.
\appendix

\onecolumngrid

\pagebreak

\vspace{10pt}

\begin{center}
    \large{\textbf{End Matter}}
\end{center}

\vspace{10pt}

\twocolumngrid

% \appendix

\section{\label{App:BodyLotion} Parallel‐superposition rheometry of the body lotion}

\begin{figure}[h]
    \centering
    \includegraphics[scale=1.0]{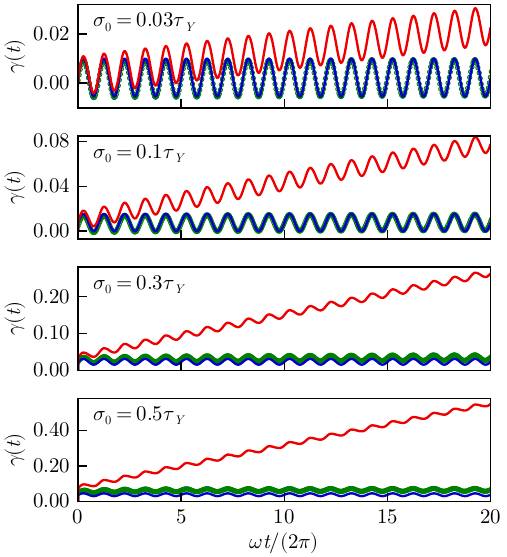}
    \caption{Comparison of parallel superposition rheometry data (green circles) with full numerical predictions from the SHB (blue) and KDR (red) models over the first 20 cycles for the body lotion. The imposed stress is $\sigma(t)=\sigma_{0}+\epsilon\sin(\omega t)$, with $\sigma_{0}/\tau_Y\in \{0.03,\,0.1,\,0.3,\,0.5\}$ (top to bottom), $\epsilon=0.1 \tau_Y$ and $\omega=1~\mathrm{rad \, s^{-1}}$.}
    \label{fig:Exp_SHB_KDR_bodylotion}
\end{figure}

\begin{figure}[h]
    \centering
    \includegraphics[scale=1.0]{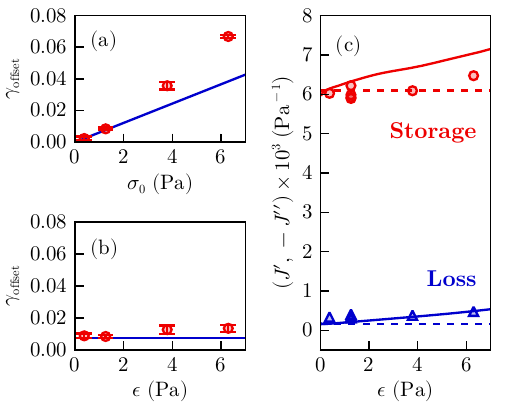}
    \caption{(a, b) Comparison between experimental strain offsets (red circles) and predictions from the SHB model (blue line) in parallel superposition tests on the body lotion. Error bars indicate the standard error across independent repetitions. (a) fixed oscillatory amplitude $\epsilon = 0.1\tau_Y$; (b) fixed mean stress $\sigma_0 = 0.1\tau_Y$. (c) Storage (red) and loss (blue) compliances as functions of $\epsilon$ for the body lotion. Markers denote experimental values from parallel superposition rheometry; solid lines correspond to conventional oscillatory shear measurements; dashed lines show predictions from the SHB model. At $\epsilon=0.1 \tau_Y$ there are four overlapping markers for values of $\sigma_0 = [0.03, \, 0.1, \, 0.3, \, 0.5]\tau_Y$. The remaining points were measured for $\sigma_0 = 0.1\tau_Y$. For all measurements $\omega=1~\mathrm{rad \, s^{-1}}$.}
    \label{fig:OffsetCompliance_bodylotion}
\end{figure}

Here we report additional parallel superposition data for the commercial body lotion to complement the main-text results. Figure \ref{fig:Exp_SHB_KDR_bodylotion} compares the measured strain response over the first 20 cycles with full numerical predictions of the SHB and KDR models. For all cases shown, the experimental response remains bounded with no discernible secular drift of the mean strain over the window displayed. The SHB prediction captures the bounded, bias-dependent waveform well, whereas the KDR prediction deviates in its mean-trend behaviour under the same loading. Figure \ref{fig:OffsetCompliance_bodylotion} quantifies the strain offsets and small-amplitude compliances extracted from all the measurements. Overall, these results reinforce the main conclusion for the body lotion: under sub-yield bias, the response exhibits recoverable deformation with bounded offsets and no measurable plasticity-induced drift, in line with SHB and inconsistent with the mean-drift behaviour predicted by KDR under similar loading protocols.

\section{\label{App:Plate-plate}Notes on stress control in parallel-disk rheometry}

In parallel-disk measurements, the rheometer controls the torque $M$ on the top plate, rather than setting a spatially uniform shear stress in the whole fluid domain. In cylindrical coordinates $(r,\theta,z)$ the measured torque is $M(t)=2\pi\int_{0}^{R}\sigma_{\theta z}(r,t)\,r^{2}\,dr$, while the kinematics are $\dot{\gamma}(r,t)=\Omega(t)\,r/h$, with $R$ the plate radius, $h$ the gap, and $\Omega$ the angular velocity. To compare experiments with the SHB and KDR models we integrate the constitutive response over $r$ when predicting $M(t)$. For compact reporting we define geometry-specific effective quantities $\sigma_{\mathrm{ref}}=4M/(3\pi R^{3})$ and $\dot{\gamma}_{\mathrm{ref}}=2\Omega R/(3h)$, chosen so that a Newtonian liquid of viscosity $\mu$ satisfies $\mu=\sigma_{\mathrm{ref}}/\dot{\gamma}_{\mathrm{ref}}$. The maximum local shear stress occurs at the rim; for the Newtonian case $\sigma_{z\theta}(R)=3\sigma_{\mathrm{ref}}/2$, so any stress control expressed via $\sigma_{\mathrm{ref}}(t)$ implies a $3/2$ amplification at $r=R$. Consequently, a sample remains entirely unyielded if the peak effective stress never exceeds $2\tau_Y/3$. For a mean-plus-oscillation protocol in parallel superposition rheometry with effective mean $\sigma_0$ and amplitude $\epsilon$, the fully unyielded condition is $\sigma_0+\epsilon\le 2\tau_Y/3$, which guarantees $\sigma_{\theta z}(r,t)\le\tau_Y$ for all $0\le r\le R$ and all times.

Using the manufacturer-reported torque (0.5~nN$\cdot$m) and angular-deflection (50~nrad) resolutions for the rheometer the and plate geometry used in this work \cite{AntonPaarMCR302}, the implied shear-stress resolution is $1.36 \times 10^{-5}$~Pa. The strain resolution is $8.33 \times 10^{-7}$ for $h=1$~mm and $4.17 \times 10^{-7}$ for $h=2$~mm. The drift-rate detection floor over a 471~s test is $1.77 \times 10^{-9}$~s$^{-1}$ for $h=1$~mm and $8.84 \times 10^{-10}$~s$^{-1}$ for $h=2$~mm. All reported measurements lie well above these detection limits.

\section{\label{App:PSP}Parallel Superposition Rheometry}

Parallel-superposition measurements were performed using the same rheometer setup with crosshatched parallel plates described in Sec.~S1 of the \emph{Supplemental Material}~\cite{SM}. 
The material samples were placed at the centre of the lower plate and the upper plate brought to the target gap using a controlled gap-closure profile with a low normal-force threshold to avoid pre-shearing. Excess material was systematically trimmed with the upper plate locked to prevent any inadvertent motion prior to testing. The experimental protocol involved superimposing a small-amplitude oscillatory stress signal onto a constant offset stress. For each measurement, the oscillatory component was applied at a fixed angular frequency of $\omega = 1$~rad/s. The constant stress component was systematically varied from 3\% to 50\% of the material’s yield stress, previously determined from steady shear rheometry (see Fig.~S1 in the \emph{Supplemental Material}~\cite{SM}). This range was selected to probe behaviour near the yield threshold while remaining safely within the unyielded  ($\leq 90$\% of the yielding point). The amplitude of the oscillatory stress was also varied in the same range. All measurements were performed under stress-controlled conditions, and the strain response was recorded over 75 oscillation cycles to ensure that a periodic steady state was reached. A relaxation step after loading was omitted, as it was not found to improve the analysis. This was due to the strain plateau at zero imposed shear stress being small and randomly positive or negative and therefore accounted for by averaging over repeats. All samples were tested at a controlled temperature of 25~$^\circ$C and were allowed to equilibrate under quiescent conditions prior to testing. A fresh sample was used for each test, even when conducted under identical conditions. All test conditions were performed in triplicate to ensure reproducibility.

\section{\label{App:Slip}Correction Protocol for Residual Wall-Slip}

Here we provide evidence that the small linear drift observed in the parallel superposition experiments is due to residual wall slip that persists despite surface roughening, and we show how we subtract it from the measured strain response. Previous work~\cite{Meeker2004} reported a quadratic dependence of slip velocity on stress in soft particle pastes, attributed to elastohydrodynamic lubrication effects in smooth walls. However, in the present work we assume that the use of roughened surfaces suppresses or significantly minimises this mechanism and that any residual slip is due to the formation of a thin depletion layer of continuous phase near the wall. This results in a lubricating film of thickness $O(10)$~nm at the top and bottom plates while the bulk moves in the middle as a solid \cite{Zhang2017}. Additionally, a microstructural mismatch between the gel/emulsion occlusions and the wall texture may result in the material not fully interlocking with the wall. This residual slip will then induce an additional apparent strain rate in the rheometer response, $\dot{\gamma}_\mathrm{slip}$, which, as shown in Ref.~\cite{Zhang2017}, can be modelled as a simple Navier slip condition. Specifically, the slip-induced strain rate is taken to be $\dot{\gamma}_\mathrm{slip} = \beta \sigma_{xy}$, where $\beta$ is a phenomenological slip parameter with units of inverse viscosity, which should be sensitive to the gap height between the two parallel plates as $\beta \sim 1/h$. Integrating this expression over time yields the cumulative slip strain $\gamma_\mathrm{slip} = \int_0^t \beta \sigma_{xy}(t') \, \mathrm{d}t'$, which, for a stress signal of the form $\sigma_{xy} = \sigma_0 + \epsilon \sin(\omega t)$, leads to the analytical expression
$\gamma_\mathrm{slip} = \sigma_0 \beta t - \epsilon \beta/\omega \left( \cos(\omega t) - 1 \right)$.

This expression can be used to correct the experimentally obtained shear strain curves.

Parallel superposition tests for the Carbopol gel were conducted at gap heights of 0.5, 1, and 2~mm. The corresponding slip parameters, extracted from these experiments, are presented in the inset of Fig.~\ref{fig:slip_factor}. The results reveal an inverse proportionality between the slip parameter and gap height, consistent with the predictions of a simple Navier slip model. This observation provides strong evidence that the slight drift in strain observed during parallel superposition is not intrinsic to the material’s viscoplastic behaviour, but rather arises as an artefact of wall slip.

All strain curves in presented in this work have been adjusted by subtracting the analytically derived slip contribution from the raw data (see Fig.~S2 and S3 in in the \emph{Supplemental Material}~\cite{SM}). A single value for the slip parameter $\beta$ was obtained by fitting the linear drift for all data at varying values of $\sigma_0$ and $\epsilon$, simultaneously. We obtained $\beta = 1.60 \times 10^{-6} \; \mathrm{Pa}^{-1} \, \mathrm{s}^{-1}$ for the Carbopol gel and $\beta = 1.76 \times 10^{-6} \; \mathrm{Pa}^{-1} \, \mathrm{s}^{-1}$ for the body lotion. This further confirms that the magnitude of the drift is determined entirely by the constant component of the applied stress and is independent of the amplitude of the oscillatory stress component, which contributes only to a small phase adjustment and offset. After correction, the strain response at longer times for all tested conditions becomes purely oscillatory and symmetric about a constant strain component, as expected for a viscoelastic solid subjected to a parallel superposition of stresses.

\end{document}